\documentclass[journal,11pt,draftcls,peerreview,onecolumn]{IEEEtran}
\usepackage{amsfonts}
\usepackage{times}
\usepackage[pdftex]{graphicx}
\usepackage{latexsym}
\usepackage{dsfont}
\usepackage{amssymb}
\usepackage{amsmath}
\usepackage{cite}
\usepackage{verbatim}
\usepackage{subfigure}

\def\bb0{{\mathbb{0}}}

\def\bb{{\mathbf{b}}}

\def\bn{{\mathbf{n}}}

\def\bs{{\mathbf{s}}}

\def\bv{{\mathbf{v}}}

\def\by{{\mathbf{y}}}
\def\bz{{\mathbf{z}}}
\def\b0{{\mathbf{0}}}

\def\bA{{\mathbf{A}}}

\def\bE{{\mathbf{E}}}
\def\bF{{\mathbf{F}}}

\def\bH{{\mathbf{H}}}
\def\bI{{\mathbf{I}}}

\def\bQ{{\mathbf{Q}}}
\def\bR{{\mathbf{R}}}

\def\bU{{\mathbf{U}}}

\def\sf0{{\mathsf{0}}}

\usepackage{amsmath}
\usepackage{amsfonts}
\usepackage{amssymb}
\usepackage{graphicx}
\usepackage{subfigure}
\usepackage{setspace}
\usepackage[nolists,nomarkers]{endfloat}

\begin{document}
%
\title{User Partitioning for Less Overhead\\in MIMO Interference Channels}
%
%
\author{
\IEEEauthorblockN{Steven~W.~Peters~and~Robert~W.~Heath,~Jr.\\}
\thanks{The authors are with The University of Texas at Austin,
1 University Station C0806, Austin, TX, 78712-0240,
Email: \{speters,rheath\}@mail.utexas.edu}
\thanks{This work was supported by a gift from Huawei Technologies, Inc.}}%

\setcounter{page}{1}
\maketitle

\begin{abstract}
This paper presents a study on multiple-antenna interference channels, accounting for general overhead as a function of the number of users and antennas in the network.
The model includes both perfect and imperfect channel state information based on channel estimation in the presence of noise.
Three low-complexity methods are proposed for reducing the impact of overhead in the sum network throughput by partitioning users into orthogonal groups. 
The first method allocates spectrum to the groups equally, creating an imbalance in the
sum rate of each group. The second proposed method allocates spectrum unequally among the groups to provide rate fairness. Finally, geographic grouping is proposed for cases where
some receivers do not observe significant interference from other transmitters. For each partitioning method, the optimal solution not only requires a brute force
search over all possible partitions, but also requires full channel state information, thereby defeating the purpose of partitioning. We therefore propose greedy methods
to solve the problems, requiring no instantaneous channel knowledge.
Simulations show that the proposed greedy methods switch from time-division to interference alignment
as the coherence time of the channel increases, and have a small loss relative to optimal partitioning only at moderate coherence times.
\end{abstract}



\newcommand{\sr}[1]{$\mathcal{#1}$}
\newcommand{\vecnorm}[2]{\left|\tilde{\bf H}_{#1}^{#2}\right|}

\section{Introduction}
Interference channels model the case of simultaneous point-to-point transmission by two or more transmitters
that do not have mutual knowledge of transmitted data for the purposes of coordinated precoding.
%
Recent work on interference channels has shown that, theoretically, the capacity of such networks increases linearly with the number of 
transmit/receive pairs in the network~\cite{CadJaf:Interference-Alignment-and-Degrees:08,GhaMotKha:Interference-Alignment-for-the:09}. 
In particular, by intelligently precoding the transmitted symbols, all the interference can be forced into a subspace of
the received space at all receivers simultaneously. This precoding operation is called interference alignment (IA).
Although IA can achieve a linear rate scaling with the number of users in a network, achieving the optimal scaling requires
network channel state information (CSI) when designing the precoders. In particular, with only two users, previous work 
has shown a loss in capacity scaling with signal-to-noise ratio (SNR) when channel coefficients are not known at the 
transmitters~\cite{ZhuGuo:Isotropic-MIMO-Interference:09,HuaJafSha:On-degrees-of-freedom-region:09}.
Other work has studied IA with statistical channel state information~\cite{Jaf:Exploiting-channel-correlations:10} or 
for other channel models~\cite{GouWanJaf:Aiming-perfectly-in-the-Dark:10}.
Iterative algorithms have been proposed that can run in a distributed fashion requiring only local channel state information at each 
node~\cite{GomCadJaf:Approaching-the-Capacity-of-Wireless:08,PetHea:Interference-Alignment-Via-Alternating:09,PetHea:Cooperative-Algorithms-for-MIMO:09}. Such algorithms
trade feedback overhead for the overhead of iterating over the wireless medium.
Previous work has shown that the number of total feedback bits for interference alignment scales as the square of the number of users in the 
network~\cite{ThuBol:Interference-alignment-with:09}. This is because the total number of wireless links grows with the square of the number of users in the network.
CSI at the transmitter can also be obtained through reciprocity, which requires calibration~\cite{GuiSloKno:A-practical-method-for-wireless:05}. 
Such a procedure trades feedback overhead for calibration and extra training overhead.

Beyond the requirement of CSI when designing precoders, there is no prior work analyzing the interference channel without channel state knowledge at the receivers.
All current methods for maximizing degrees of freedom (DOF, the pre-log factor in the sum capacity term related to the total number of spatial streams in the network) 
for the interference channel require channel training and estimation at each node even if no feedback mechanism is employed.
The requirement of CSI, even if only at the receivers, may still dominate communication in an interference channel with many users, since
training is known to effectively reduce the degrees of freedom of a point-to-point link~\cite{HasHoc:How-much-training-is-needed:03}.
With low-to-moderate coherence times, the training required to estimate the $K^2$ wireless channels in a $K$-user MIMO interference channel can 
last nearly as long as the coherence time, leaving a very short amount of time for IA transmission before the CSI becomes stale.
Time and frequency synchronization among all nodes is also required for interference alignment adding to the overhead burden of the network.

To mitigate the domination of scaling overhead in large interference channels, prior work has considered clustering in a cellular network
based on spatial proximity~\cite{TreGui:Clustered-Interference-Alignment:09,TreGuiRie:A-clustered-alignment-based-interference:09},
but this clustering is done without optimization and does not consider overhead.
Others have considered the impact of imperfect CSI on the achievable sum rate of interference 
alignment~\cite{TreGui:Cellular-interference-alignment:09}, but considered only the case where all links have the same channel estimation error.
The number of bits of limited feedback desired for single-antenna
interference alignment was investigated in~\cite{ThuBol:Interference-alignment-with:09}. Overhead due to training was neglected in both cases. 

Interference alignment-type transmitters with no transmit CSI have recently been 
proposed~\cite{GouWanJaf:Aiming-perfectly-in-the-Dark:10,HuaJafSha:On-degrees-of-freedom-region:09}, resulting in reduced network degrees of freedom.
Such work makes the assumption that the network is operating in an environment where training and feedback overhead will dominate, and the total IA throughput
will be smaller than a suboptimal strategy with no feedback. This assumption is valid in quickly varying channels. The overhead conditions are not quantified.
This paper makes no such assumption and instead addresses the question, ``how much overhead makes IA infeasible?''  The question has not 
been addressed in the literature, and its answer is unclear. With very static channels, 
we can dedicate long training sequences to generate high-fidelity training estimates that will be accurate for a long period of time.
Further, with quickly varying channels, obtaining a large amount of channel information is infeasible. For all the cases between these
two extremes, the overhead must be quantified.

In this paper we account for overhead in MIMO interference channels through an overhead penalty factor on the sum throughput.
The model assumes synchronized narrowband block fading with overhead requiring access to the wireless medium at the beginning of each frame. Using this model we 
show that the 
achieved sum rate with overhead of interference alignment will go to zero with a large number of users, even if the only overhead in the network is due to training. 
That is, even with a minimal amount of overhead (minimum training lengths, no feedback, no synchronization overhead, no medium access control overhead, etc.), 
IA does not have asymptotically increasing sum rate as the number of users grows large.
We then show that, if the overhead grows faster than linearly with the number of users in the network, partitioning the network into orthogonally transmitting
groups can increase the effective degrees of freedom. The rest of the paper is devoted to developing smart partitioning methods.

First, we consider a connected interference channel, where spatial clustering is ineffective because of the proximity of all users. 
We derive an optimization to maximize the sum rate when each group is allocated an equal amount of transmission time, 
and the solution to this optimization is shown to be too complex to serve its purpose, 
requiring global channel state information and comprehensive search. 
We therefore propose a greedy algorithm that requires only large scale information (i.e., channel magnitude) on the link between each
transmit/receive pair, but not for the interfering links. The availability of such information is justified because it is likely to be correlated across channel
realizations. Based on an approximation to the sum rate for interference alignment using linear precoding,
the proposed algorithm efficiently partitions the network into IA groups.
Relative to our previous work~\cite{PetHea:Orthogonalization-to-reduce-overhead:10}, this paper introduces new partitioning algorithms, proposes geographical and
equal-rate grouping, and includes analysis on training length.

Second, we derive an equal-rate unequal-time allocation between groups to enforce sum-rate-fairness rather than time-slot-length fairness. This algorithm is shown to require
a small modification to the equal-time allocation algorithm and an additional final step solving a linear system of equations. This solution is again based on a connected
interference channel where spatial clustering is not beneficial. 
In an unconnected network, grouping together users that are geographically separated may allow them to transmit
nearly orthogonal in space with higher throughput due to significant path loss from interfering transmitters. Conversely, a network can be partitioned into groups that are
nearly mutually orthogonal in space, such that the groups can transmit simultaneously (rather than the users transmitting simultaneously while groups transmit orthogonally).
Finally, we derive greedy algorithms for both of these scenarios based on position information obtained through GPS or similar positioning methods. 
The spatial clustering algorithms are well-suited for dense ad hoc 
networks~\cite{TreGui:Cellular-interference-alignment:09,TreGui:Performance-of-Interference-Alignment:10,AndShakHea:Rethinking-information-theory:08}, 
where a natural spatial clustering may not be present or is distorted because of overhead.
Assuming the existence of an IA-enabling mechanism built into the network, these algorithms require no additional overhead. 

In summary, this paper proposes a suite of transmission strategies, and a method for choosing among them, that trade increased overhead for increased
capacity, or decreased capacity for decreased overhead. The strategies presented here are parameterized by a single scalar parameter, the number of groups with which to
partition the network. The most complex strategy considered is interference alignment through the entire network; the simplest strategy considered is time division
multiple access (TDMA) across the entire network. By partitioning the network into groups that transmit mutually orthogonally, but using IA inside the groups, the gap between
IA and TDMA is filled using very little network knowledge and processing. Previous work on grouping, for instance for network 
MIMO~\cite{BocHuaAle:Network-MIMO-with:08} and interference alignment~\cite{TreGui:Clustered-Interference-Alignment:09}, was performed with the 
overall goal of trading overhead and rate without explicitly taking overhead into account. Previous efforts to reduce the overhead of 
IA transmissions, including~\cite{GomCadJaf:Approaching-the-Capacity-of-Wireless:08,ThuBol:Interference-alignment-with:09}, assume that all the users
are using IA simultaneously, which this paper shows is often suboptimal. Finally, previous work on imperfect channel state information in interference
channels finds rate bounds but does not optimize these rates as a function of length of the training, as this paper studies.

This paper is organized as follows: Section~\ref{sec:model} presents the model utilized in this paper; Section~\ref{sec:partitioning} discusses the problem of partitioning in general
and shows why optimal partitioning is impractical; Section~\ref{sec:greedy} proposes greedy algorithms for partitioning the network based on equal time allocation, equal sum rate
allocation, and geographic nearness; Section~\ref{subsec:imperfect_csi} analyzes the relationship between partitioning and training overhead; Section~\ref{sec:sims} presents computational simulations while Section~\ref{sec:conclusion} concludes the paper and points toward future work.

Finally, a word on notation. The $\log$ refers to $\log_2$. Bold uppercase letters, such as ${\bf A}$, denote 
matrices, bold lowercase letters, such as ${\bf a}$, denote column vectors, and normal letters $a$ denote scalars.
The letter $\mathbb{E}$ denotes expectation, $\mathbb{C}$ is the complex field, $\max\{a,b\}$ denotes the maximum of $a$ and $b$, 
$\|\bA\|_F$ is the Frobenius norm of matrix ${\bA}$, and $\left|\bA\right|$ is the
determinant of square matrix $\bA$. The empty set is denoted as $\emptyset$, the identity matrix of appropriate dimension is $\bI$, 
and $\bI_{A\times B}$ is the $A\times B$ truncated identity matrix. 

\section{System Model}\label{sec:model}
\begin{figure}
\centering
\includegraphics[width=3.5in]{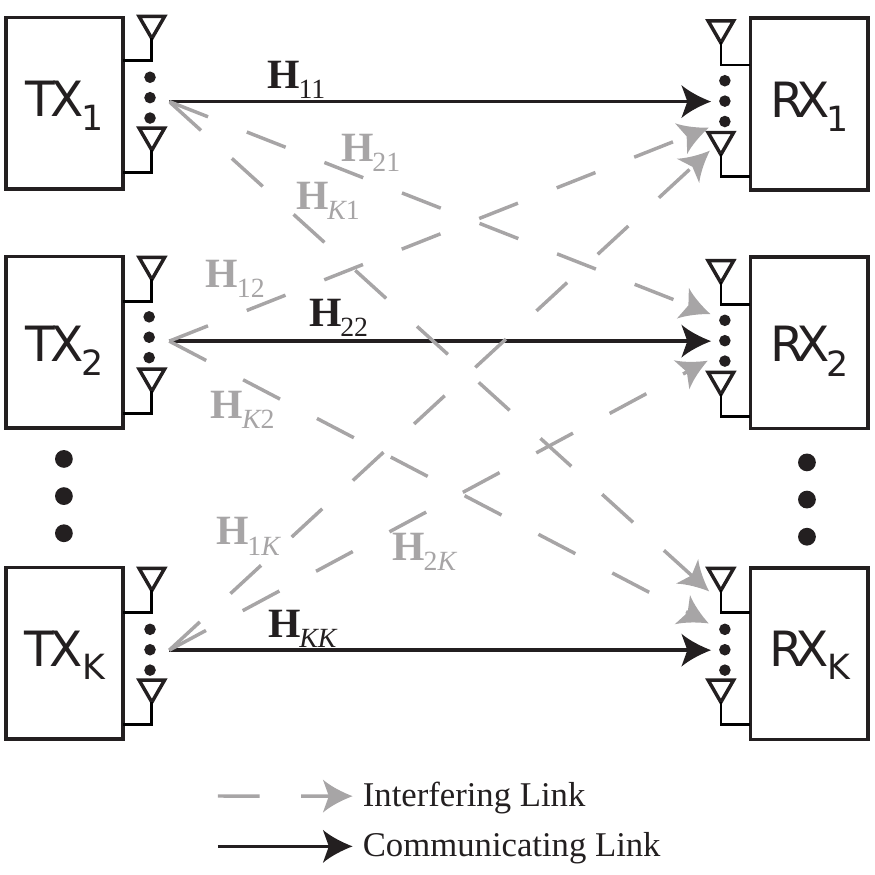}
\caption{The MIMO interference channel. Each transmitter is paired with a single receiver. In the model considered in this paper,
the channels $\bH_{k,\ell}$ are block fading with coherence time $T_{k,\ell}$.}
\label{fig:channel_model}
\end{figure}

We consider a distributed MIMO network with $2K$ nodes. $K$ of the nodes have data to transmit via their $N_t$ antennas
to the other $K$ nodes, each with $N_r$ antennas, with no multicasting or cooperative transmission. Transmitter $k\in\{1,\dots,K\}$ has data
destined only for receiver $k$. We assume a narrowband block fading model where the $N_r\times N_t$ matrix channel
$\bH_{k,\ell}$ between transmitter $\ell\in\{1,\dots,K\}$ and receiver $k\in\{1,\dots,K\}$ is independently generated every $T$ symbol periods
$\forall k,\ell$. 
We assume transmissions are frame and frequency synchronous.
Thus, at any fixed moment in time, there is a $K$-user MIMO interference channel with $N_t$ antennas at each transmitter and $N_r$ antennas at each receiver, as illustrated in 
Figure~\ref{fig:channel_model}. 
We consider scenarios where interference alignment is considered to be theoretically amenable; that is, we consider channels in which, without overhead, 
IA would be a good candidate transmission strategy, with strong channels between all nodes. 
In scenarios where IA is not desirable, such as when interference is much stronger than the signal, 
receiver methods such as successive interference cancellation (SIC) may be more attractive~\cite{And:Interference-cancellation-for-cellular-systems:05}.
The assumption that all nodes have identical coherence times is justified because of previous work showing that multiuser transmission is severely degraded in quickly changing
channels~\cite{Cai:MIMO-downlink-joint:06,ZhaHeaKou:Mode-switching-for-the-MIMO:09}, 
meaning all candidates for interference alignment are likely to have relatively static channels. Analysis for different coherence times for each link is left for future 
work.

\begin{figure}
\centering
\includegraphics[width=5in]{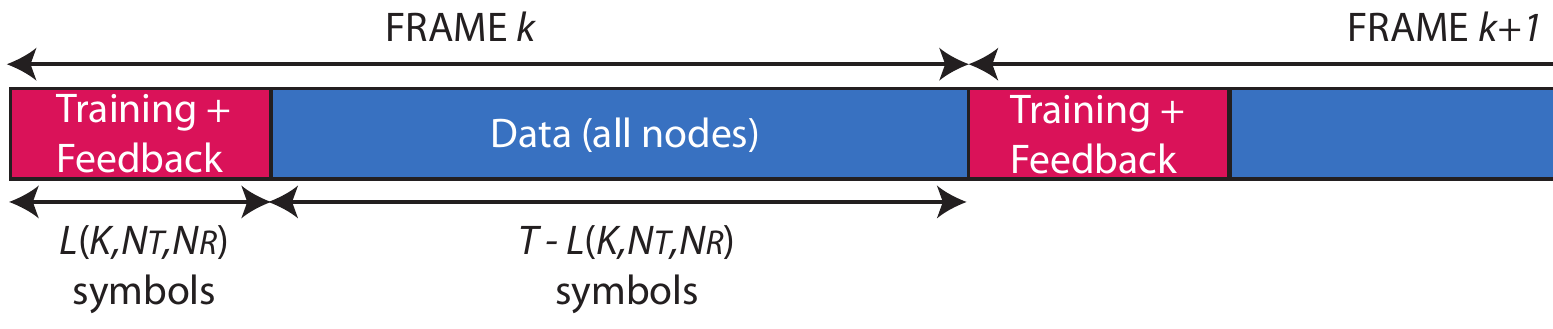}
\caption{Illustration of the communication frame used for the model in this paper. The beginning of the frame is used for overhead of any nature, consuming $\mathcal{L}(K,N_t,N_r)$ 
symbols. The remaining $T-\mathcal{L}(K,N_t,N_r)$ symbols are used for data transmission. New channels, independent of previous realizations, are generated at the end of the frame.}
\label{fig:frames}
\end{figure}

Communication is divided into frames of period $T$ symbols, as shown in Figure~\ref{fig:frames}. The beginning of each frame is devoted to overhead, which may include 
training, feedback, synchronization, higher layer overhead, and so on. We do not make assumptions about the source or amount of overhead.
Later we will explicitly model channel training and estimation, but this will not preclude the existence of other overhead sources.
For channel estimation, the transmitters send mutually orthogonal training
sequences since the network is connected (i.e., spatially dense). This training is necessary not only for coherent detection but also for CSI feedback
required to exploit the full degrees of freedom in the 
network~\cite{ZhuGuo:Isotropic-MIMO-Interference:09,HuaJafSha:On-degrees-of-freedom-region:09}. 
Although reciprocity can be exploited~\cite{CadJaf:Interference-Alignment-and-Degrees:08}, it requires double the training and a special calibration procedure among all the nodes
in the network~\cite{GuiSloKno:A-practical-method-for-wireless:05}. 
Overhead time is $\mathcal{L}(K,N_t,N_r)\le T$ symbol periods. 
Thus overhead requires a fraction $\alpha=\min\{\mathcal{L}(K,N_t,N_r)/T,1\}$ of the frame, while data is transmitted during the remaining $\overline{\alpha}=1-\alpha$. This overhead model is an extention of the model in~\cite{HasHoc:How-much-training-is-needed:03} to the MIMO interference
channel.

The data transmission portion of the frame begins after the first $\mathcal{L}(K,N_t,N_r)$ symbols and ends when the channel changes $T$ transmissions later. 
Information theoretic results, which neglect overhead, suggest that all transmitters should send aligned signals simultaneously to achieve the maximum degrees
of freedom in the channel and thus approach its sum capacity with high transmit 
power~\cite{CadJaf:Interference-Alignment-and-Degrees:08,GhaMotKha:Interference-Alignment-for-the:09}.
The overhead portion of the frame has given the transmitters sufficient information to design linear precoders. While linear precoding may not be 
sum-rate-optimal~\cite{GhaMotKha:Interference-Alignment-for-the:09},
it is a practical approach for immediate implementation because of the simplicity of the receiver signal processing.
Transmitter $\ell$ sends $S_\ell$ spatial streams to receiver $\ell$. At symbol period $n$, the signal observed by receiver $k\in\{1,\dots,K\}$ is 
\begin{equation}
\by_k[n] = \sqrt{\rho_{k,k}}\bH_{k,k}\bF_k\bs_k[n] + \sum_{\substack{\ell=1\\ \ell\ne k}}^K\sqrt{\rho_{k,\ell}}\bH_{k,\ell}\bF_\ell\bs_\ell[n] + \bv_k[n],
\end{equation}
where $\rho_{k,\ell}=E_k\gamma_{k,\ell}$, $E_k$ is the transmit power from transmitter $k$, $\gamma_{k,\ell}$ is the fading coefficient from transmitter $k$ to receiver $\ell$,
$\bH_{k,\ell}$ is the $N_r\times N_t$ MIMO channel from transmitter $k$ to receiver $\ell$,
$\bF_\ell$ is the $N_t\times S_\ell$ unit-norm linear precoder used at transmitter $\ell$, $\bs_\ell$ is the $S_\ell\times 1$ vector 
of symbols sent by transmitter $\ell$, and $\bv_k$ is zero-mean white circularly symmetric zero-mean
complex Gaussian noise with covariance matrix $\mathbb{E}\bv_k\bv_k^*=\bR_k$.
The rest of the paper examines the implications of overhead as a function of the number of users and proposes methods to find a balance between overhead and capacity gains.

\section{Optimal Partitioning to Reduce Overhead}\label{sec:partitioning}
This section introduces and motivates the notion of network partitioning to reduce overhead. 
We first consider the case of maximizing the sum rate of a network with perfect channel estimation. 
The model described in Section~\ref{sec:model} is a $K$-user MIMO interference channel during the data portion of the frame.
Assuming the training performed in the first part of the frame results in perfect CSI at both transmitter and
receiver, with the overhead model described in Section~\ref{sec:model} and maximum likelihood reception, 
the sum rate of the network in bits per transmission for a particular frame is then
\begin{equation}
R_{\rm sum} = \overline{\alpha}\sum_{k=1}^K\log\biggl|\bI + \biggl(\bR_k +
  \sum_{\ell\ne k}^K\rho_{k,\ell}\bH_{k,\ell}\bF_\ell\bF_\ell^*\bH_{k,\ell}^*\biggr)^{-1}\rho_{k,k}\bH_{k,k}\bF_k\bF_k^*\bH_{k,k}^*\biggr|.
\label{eq:sum_rate}
\end{equation}

When all the transmitters are communicating during the data portion of the frame, the effective throughput is reduced by a factor of $\overline{\alpha}$
relative to the information-theoretic sum rate. The reduction factor $\overline{\alpha}$ is a function of the number of symbols required for overhead and the coherence time of the channel.
Overhead includes symbols required for training, feedback, synchronization, or any other spectrum utilization not used for communication of data.
It is thus a function of the number of users in the channel and the number of antennas at each node. 

\begin{figure}
\centering
\includegraphics[width=3.5in]{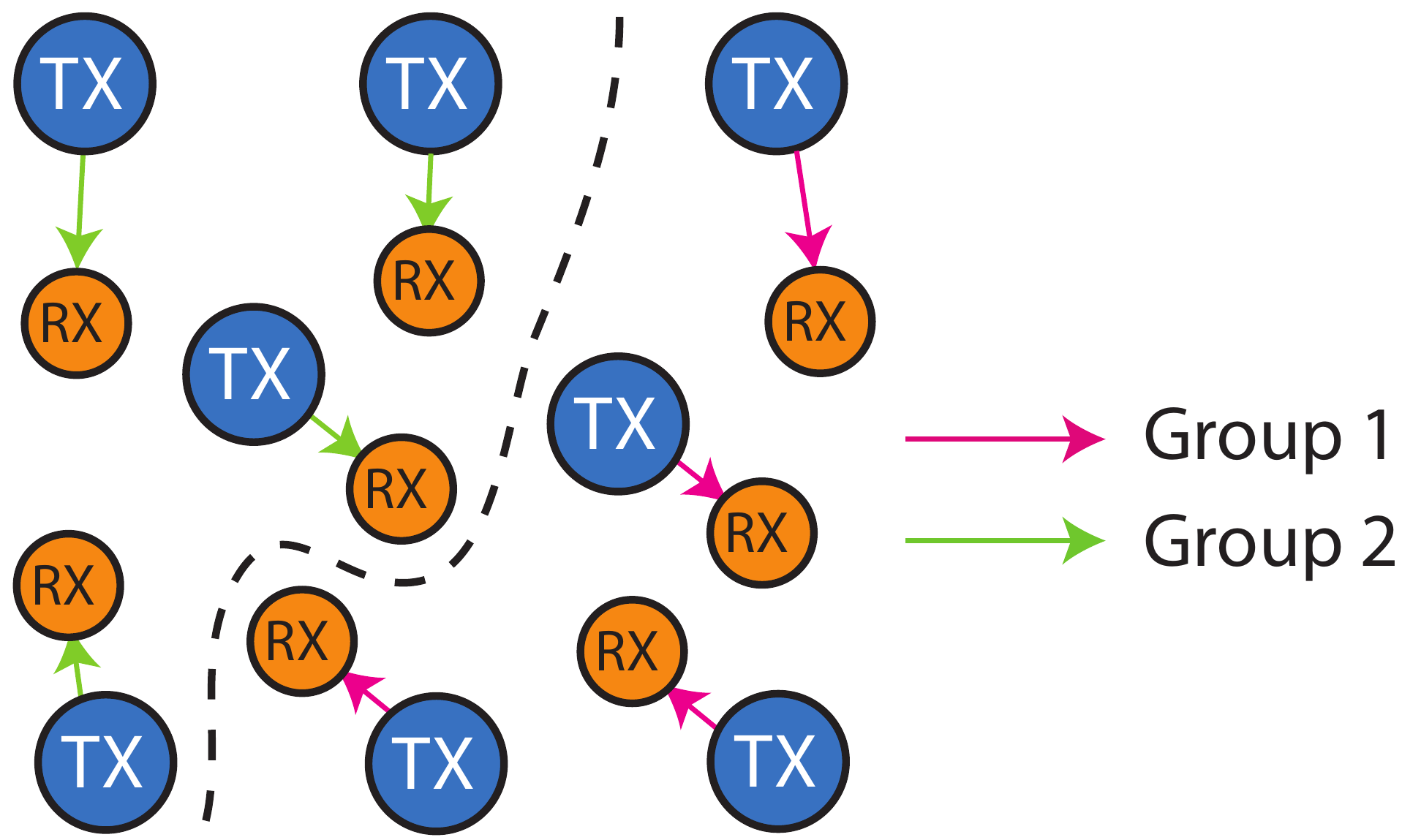}
\caption{Illustration of a partition of the $K$-user interference channel into two $K/2$-user interference channels transmitting orthogonally to each other.}
\label{fig:splitting}
\end{figure}

Our claim is that, if the overhead in the network scales faster than linearly with the number of users in the network, then the sum rate of the network may be
increased through partitioning. 
Figure~\ref{fig:splitting} illustrates the concept of partitioning.
Instead of all the transmitters sending simultaneously throughout the data portion of the frame, the frame is divided into $P$ sub-frames, each with an overhead 
and data portion. 
If overhead does scale faster than linearly with $K$,
then splitting the interference channel into $P$ equally-sized interference channels utilizing the spectrum equally but orthogonally will reduce overhead. That is, if $P>1$,
\begin{eqnarray}
P\left(\frac{T/P-\mathcal{L}(K/P,N_t,N_r)}{T}\right) & = & \frac{T-P\mathcal{L}(K/P,N_t,N_r)}{T}\nonumber\\
{} & > & \frac{T-\mathcal{L}(K,N_t,N_r)}{T}.
\end{eqnarray}
Previous work has shown that feedback overhead for IA 
scales with the square of the number of users~\cite{ThuBol:Interference-alignment-with:09}. A measurement study of a network not even performing coordinated transmissions
found that overhead scaled faster than linearly with the number of users~\cite{HenSpaKim:A-wireless-interface-type:03}. 
Orthogonalization thus has significant potential to improve the effective sum rate by reducing total network overhead. 

\begin{figure}
\centering
\includegraphics[width=3.5in]{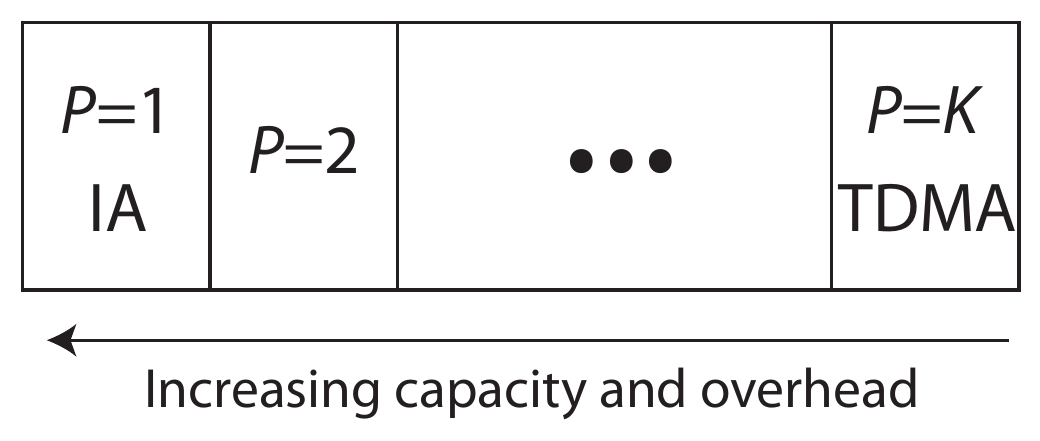}
\caption{Illustration of the parameterized suite of transmission strategies partitioning provides. With $P=1$, all $K$ users transmit simultaneously using interference
alignment, which provides capacity gains at the cost of increased overhead. At $P=K$, the users transmit orthogonally in TDMA/FDMA fashion, with relatively low
complexity and overhead, but also lower capacities. For $2\le P\le K-1$, the interference channel is partitioned into smaller groups which transmit IA within the groups,
but orthogonal to other groups.}
\label{fig:spectrum}
\end{figure}

Since the capacity of IA is known to increase with the number of users
$K$~\cite{CadJaf:Interference-Alignment-and-Degrees:08,GhaMotKha:Interference-Alignment-for-the:09,YetJafKay:Feasibility-Conditions-for-Interference:09},
forcing all users to transmit orthogonally (time division multiple access, TDMA) is not optimal in general, though in some cases it may be. 
We therefore propose a suite of transmission strategies, parameterized by the number of orthogonal groups $P$, spanning complexity and overhead from interference 
alignment to TDMA, as illustrated in Figure~\ref{fig:spectrum}. That is, for $P=1$, all the users are transmitting simultaneously using IA, and with $P=K$,
the users are transmitting orthogonally in time-division fashion. For $1<P<K$, the network is using a hybrid of the two techniques.

Note that since the original $K$ users were modeled as a connected interference channel, where all
receivers observe a signal from all transmitters above the noise floor, any subset of
transmit/receive pairs, in isolation, may also be modeled as a connected interference channel. 
The interference channel can be modeled as a connected graph~\cite{Jun:Graphs-Networks-and-Algorithms:08}. A vertex $v_k$ would include both the transmitter and receiver for
user pair $k$. The cost of each edge could be the signal-to-noise ratio from the transmitter in one vertex to the receiver in the other vertex. In this model, the edge cost is assumed to be 
reciprocal, though this does not imply that the channel is reciprocal. The weight associated with each vertex is the signal-to-noise ratio from the transmitter in one vertex to the 
receiver in the same vertex. 

Graph partitioning is an important, well-studied problem in combinatorial optimization~\cite{KerLin:An-efficient-heuristic-procedure:70}. 
Standard graph partitioning methods, however, are not directly applicable to the problem
considered in this paper. The main reason is that overhead is difficult to incorporate into the graph model. That is, the sum weight of a group will depend on how many vertices are 
assigned to the group, which is not reflected in the static weight/cost model. In a non-connected interference channel, where some receivers do not observe interference from some
transmitters, graph partitioning can be directly applied to produce non-orthogonal groups that attempt to transmit IA at the same time. This is described
in more detail in Section~\ref{subsec:spatial}.
We thus develop novel methods for the partitioning desired in our network model.

If users in the interference channel are partitioned into $P$ index sets $\{\mathcal{K}_p\}$, with $K_p=|\mathcal{K}_p|$ users in the $p$th group, then the sum rate of the network becomes
\begin{equation}
\hat{R}_{\rm sum} = \sum_{p=1}^P\overline{\alpha}_p\sum_{k\in\mathcal{K}_p}\log\biggl|\bI+\biggl(\bR_k+
    \sum_{\substack{\ell\in\mathcal{K}_p\\\ell\ne k}}\rho_{k,\ell}\bH_{k,\ell}\bF_\ell\bF_\ell^*\bH_{k,\ell}^*\biggr)^{-1}\rho_{k,k}\bH_{k,k}\bF_k\bF_k^*\bH_{k,k}^*\biggr|,
\label{eq:partition_rate}
\end{equation}
where 
\begin{equation}
\overline{\alpha}_p=\frac{T/P-\mathcal{L}(K_p,N_t,N_r)}{T}.
\end{equation}
This extension of~(\ref{eq:sum_rate}) sums the rate of each point-to-point MIMO link inside each group 
($k\in\mathcal{K}_p$), and over all groups $(p\in\{1,\dots,P\})$, where only users in the same group
interfere with each other. 
We then aim to solve the following optimization:
\begin{eqnarray}
\mathrm{maximize} & \hat{R}_{\rm sum}\nonumber\\
\mathrm{with~respect~to} & P\in\mathbb{N}_1,K_p\in\mathbb{N}_1\forall p, \bF_\ell\in\mathbb{C}^{N_t\times S_\ell}\forall\ell\nonumber\\
\mathrm{subject~to} & \sum_{p=1}^PK_p=K,~\|\bF_\ell\|\le 1.
\end{eqnarray}
The solution to this optimization is computationally complex and involves not only a brute force search over every possible grouping, but also the 
calculation of the desired precoders for each grouping. Neglecting the precoder calculations, and assuming that we have a priori knowledge that the optimal 
partition is to equally distributed users across groups\footnote{This assumption is a good approximation in most cases, but is not optimal in every case. Not making this
assumption greatly increases the search complexity even further.}, the number of searches required is still~\cite{KerLin:An-efficient-heuristic-procedure:70}  
\begin{equation}
\sum_{P=1}^K\frac{1}{P!}\left(\begin{array}{c}K\\K/P\end{array}\right) \left(\begin{array}{c}K-K/P\\K/P\end{array}\right) \cdots \left(\begin{array}{c}K/P\\K/P\end{array}\right).
\label{eq:optimal_complexity}
\end{equation}
Further, such an optimization requires each link $\bH_{k,\ell}$ to be trained and estimated,
negating the overhead reduction that partitioning provides. Obviously this is not a practical way to optimize overhead in interference networks.
In the next section we present a greedy method for performing channel partitioning with only channel quality information. 

\section{Greedy Partitioning}\label{sec:greedy}
The sum-rate-optimal partition was shown at the end of Section~\ref{sec:partitioning} to be too complex for implementation. We thus turn to heuristic approaches to reduce
not only computational complexity but also the amount of network knowledge required for implementation. We first develop a greedy method of partitioning the network
where each group is allocated the same amount of time for transmission. We then develop a method for allocating time in an unbalanced fashion to make each group's sum rate equal. 
Lastly we consider geographic partitioning methods that can exploit an unconnected interference channel.

\begin{figure}
\centering
\includegraphics[width=3.5in]{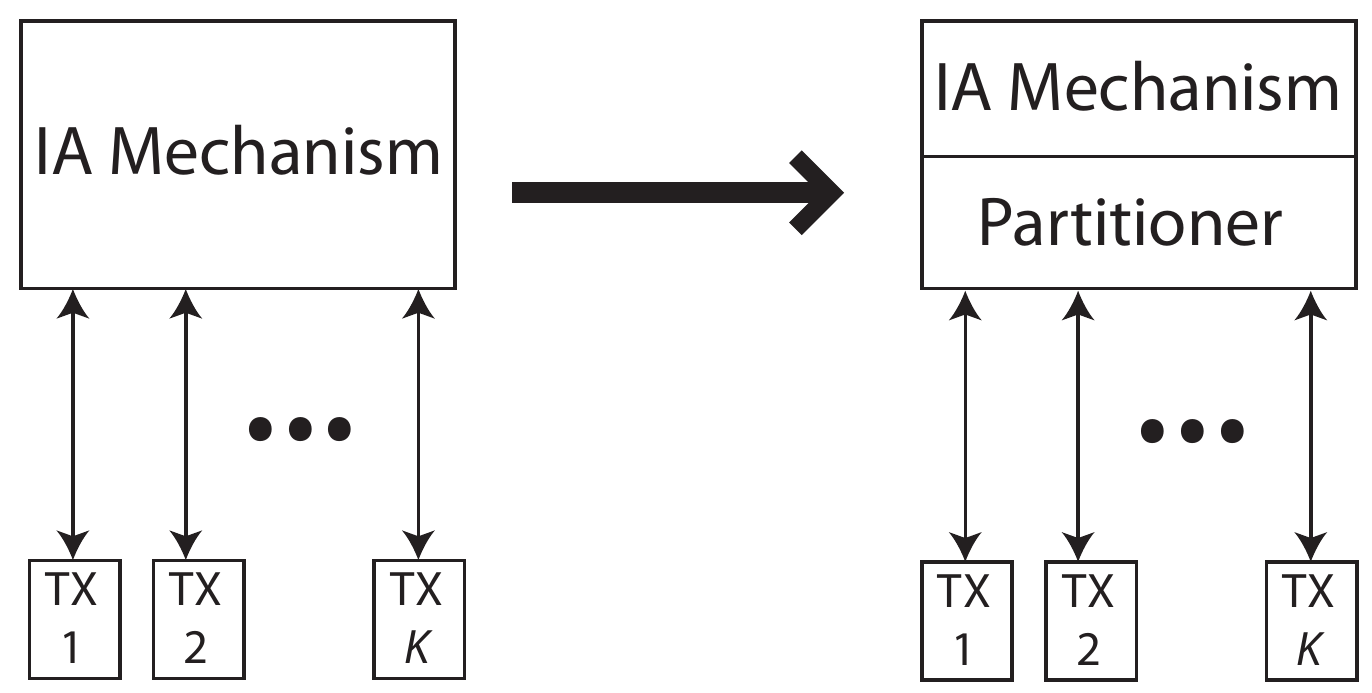}
\caption{Illustration of the modifications needed to transform an IA-only system into a partitioning system. The partitioning function, if implemented in a greedy manner
as explained in this section, requires no more communication to or from the transmitters as the IA-only system.}
\label{fig:piggy}
\end{figure}
For the following algorithms we assume a network mechanism exists to allow IA transmissions simultaneously from all transmitters if needed. 
Such a mechanism can be a central controller or a distributed protocol. The partitioning can be piggy-backed onto this mechanism, as illustrated in Figure~\ref{fig:piggy}, 
with no additional communications overhead, either through a wired backbone or the wireless medium. 

\subsection{Balanced Time Allocation}\label{subsec:max_rate}
To develop a greedy algorithm for partitioning the network, we must first define a \emph{selection function} that assigns a value of placing a user
in a group. This function would ideally be the sum rate increase of placing a user in a group. This is difficult in multiuser networks 
since the actual sum rate increase will depend on which future users are assigned to the group---knowledge that is unavailable in a greedy algorithm, which makes the locally
optimum choice at each step without global knowledge.
Instead we resort to an approximation of this sum rate increase.

\begin{figure}
\centering
\includegraphics[width=5in]{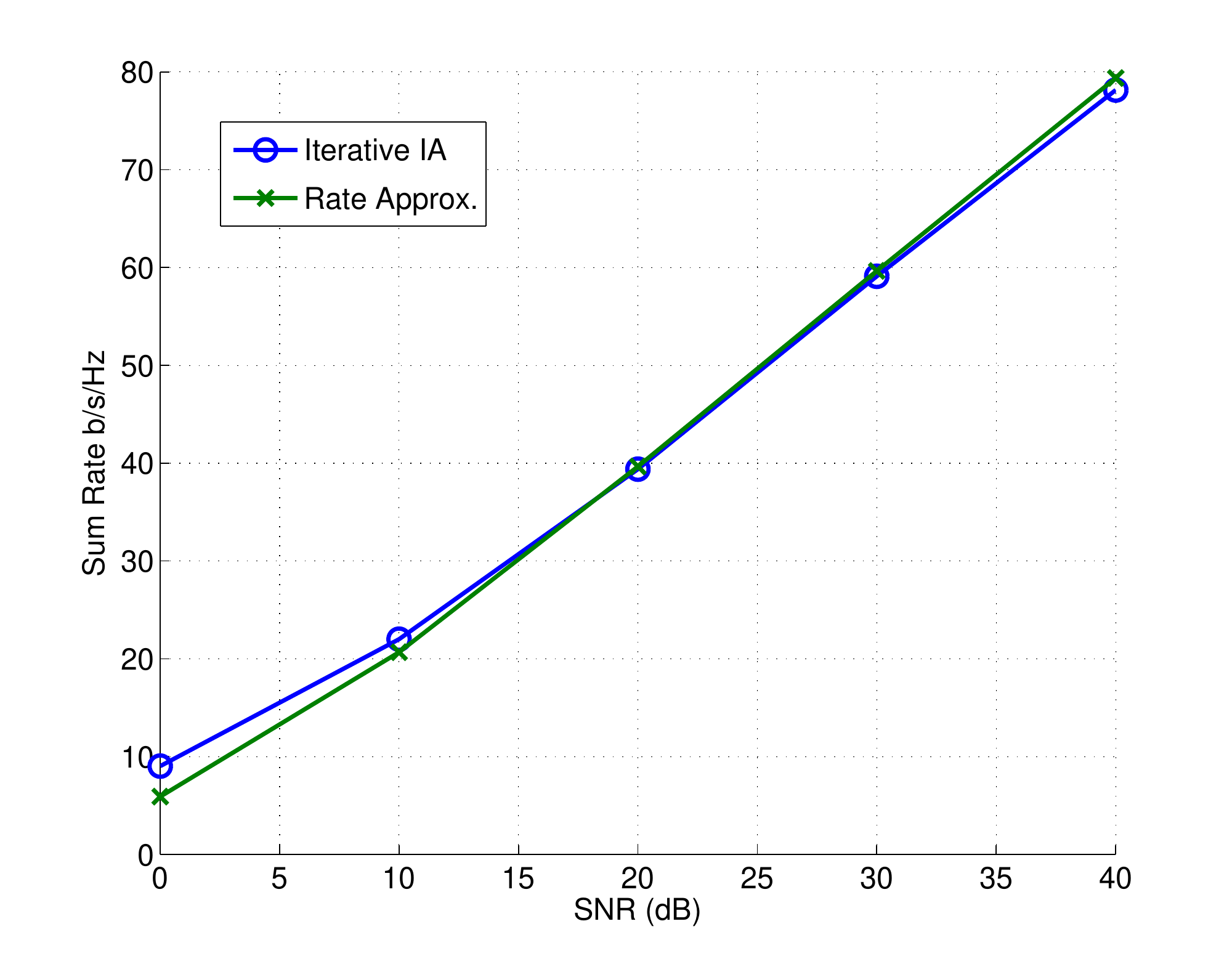}
\caption{Sum rate versus SNR of the approximation in~(\ref{eq:approx_rate}) for the 3-user MIMO interference channel with 4 antennas at each node and 2 streams per user. In this simulation and unless noted otherwise, SNR is the signal-to-noise ratio of all links in the interference channel,
including interfering links.}
\label{fig:approx}
\end{figure}

After partitioning the $K$-user interference channel into $P$ orthogonal groups, group $p$ will be a $K_p$-user interference channel that is restricted to
utilizing only $1/P$ of the spectrum or coherence interval. This enforces a time-sharing fairness constraint while attempting to maximize sum throughput
for the entire frame. An equal-rate-per-group design, which involves unbalanced time allocations, will be investigated in Section~\ref{subsec:fairness}. 
Thus, interference alignment is a reasonable choice for precoder design in each group.
Although interference alignment requires extensive CSI and calculation of precoders to find the exact sum rate, we note that the precoder solutions
are independent from the direct links $\{\bH_{k,k}\}, \forall k$. Thus, with interference alignment, the expected throughput will be approximately the rate obtained from randomly generating
orthogonal precoders $\bQ$ and combiners $\bf \Phi$ of correct rank drawn uniformly from the Grassmann manifold in the absence of interferers because
of our lack of knowledge of the channel state affecting the precoders and combiners.
We then approximate the expected rate for user $k$ in group $p$ to be
\begin{equation}
\overline{R}_{k,p} \approx \overline{\alpha}_{k,p}\mathbb{E}_{{\bf\Phi},\bQ}\log\left|\bI+\frac{\rho_{k,k}}{S_k}{\bf\Phi}^*\bH_{k,k}\bQ\bQ^*\bH_{k,k}^*{\bf\Phi}\right|,
\label{eq:rand_precoder_rate}
\end{equation}
where the scaling factor $\rho_{k,k}/S_k$ is for power normalization and $\overline{\alpha}_{k,p} = (T/P-\mathcal{L}(K_p+1,N_t,N_r))/T$. 
The expectation in~(\ref{eq:rand_precoder_rate}) is an approximation because we draw $\bQ$ 
and $\bf\Phi$ independently, whereas actual IA precoders and combiners are not mutually independent.
We then let $\bQ=\bQ_U\bI_{N_t\times S_k}$ and ${\bf\Phi}={\bf\Phi}_U\bI_{N_r\times S_k}$, where $\bQ_U$ and ${\bf\Phi}_U$ are random unitary matrices of appropriate dimension
and $\bI_{A\times B}$ is the $A\times B$ truncated identity matrix.
Defining $\hat{\bH}_{k,k}={\bf\Phi}_U^*\bH_{k,k}\bQ_U$, then
\begin{equation}
\overline{R}_{k,p} \approx \overline{\alpha}_{k,p}\mathbb{E}_{{\bf\Phi}_U,\bQ_U}\log\left|\bI+\frac{\rho_{k,k}}{S_k}\bI_{N_r\times S_k}^T\hat{\bH}_{k,k}\bI_{N_t\times S_k}\bI_{N_t\times S_k}^T
  \hat{\bH}_{k,k}^*\bI_{N_r\times S_k}\right|\label{eq:approx_1}.
\end{equation}
Then, defining the matrix $\hat{\bH}_k = \bI_{N_r\times S_k}^T\hat{\bH}_{k,k}\bI_{N_r\times S_k}$,~(\ref{eq:approx_1}) 
becomes
\begin{eqnarray}
\overline{R}_{k,p} & \approx & \overline{\alpha}_{k,p}\mathbb{E}_{{\bf\Phi}_U,\bQ_U}\log\left|\bI+\frac{\rho_{k,k}}{S_k}\hat{\bH}_{k}\hat{\bH}_{k}^*\right|\nonumber\\
{} & = & \overline{\alpha}_{k,p}\mathbb{E}_{{\bf\Phi}_U,\bQ_U}\log\left|\bI + \frac{\rho_{k,k}}{S_k}\hat{\bf\Sigma}_{k}^2\right|\nonumber\\
{} & = & \overline{\alpha}_{k,p}\mathbb{E}_{{\bf\Phi}_U,\bQ_U}\sum_{i=1}^{S_k}\log\left(1+\frac{\rho_{k,k}}{S_k}\hat{\sigma}_{k,i}^2\right),
\label{eq:approx_2}
\end{eqnarray}
where $\hat{\bf\Sigma}_k$ is the $S_k \times S_k$ diagonal matrix of singular values of $\hat{\bH}_k$
Precise calculation of~(\ref{eq:approx_2}) is not trivial, so we resort to the bound $\sum_{i=1}^{S_k}\log(1+\sigma^2_i)\le S_k\log(1+(1/S_k)\sum_{i=1}^{S_k}\sigma^2_i)$.
This bound is tight when the singular values are roughly equal. Then,~(\ref{eq:approx_2}) can be rewritten as
\begin{equation}
\overline{R}_{k,p} \approx \overline{\alpha}_{k,p}S_k\mathbb{E}_{{\bf\Phi}_U,\bQ_U}\log\left(1+\frac{\rho_{k,k}}{S_k^2}\|\hat{\bH}_k\|_F^2\right).
\label{eq:approx_3}
\end{equation}
Again with no knowledge of the channels $\{\bH_{k,\ell}\},k\ne\ell$ on which IA precoder design is based, we resort to computing the expectation
\begin{equation}
\mathbb{E}_{{\bf\Phi}_U,\bQ_U}\|\hat{\bH}_A\|_F^2 = \frac{\rho_{k,k}S_k^2}{N_tN_r}\mathbb{E}_{{\bf\Phi}_U,\bQ_U}\|\hat{\bH}_{k,k}\|_F^2 = \frac{\rho_{k,k}S_k^2}{N_tN_r}\|\bH_{k,k}\|_F^2.
\label{eq:normchan}
\end{equation}
Using Jensen's inequality~\cite{Cov:Elements-of-Information-Theory:91}, we subsitute the right side of~(\ref{eq:normchan}) into~(\ref{eq:approx_3}) and finally have
\begin{equation}
\overline{R}_{k,p} \approx \overline{\alpha}_{k,p}S_k\log\left(1+\frac{\rho_{k,k}}{N_tN_r}\|\bH_{k,k}\|_F^2\right).\label{eq:approx_rate}
\end{equation}
This approximation is justified via the plot in Figure~\ref{fig:approx} for a 3-user 4-antenna system transmitting 2 streams per user. Despite the seemingly large number of approximations 
made in the derivation, the estimate is surprisingly tight, especially at moderate-to-high SNR.

The estimation of~(\ref{eq:approx_rate}) requires $K_p$, $N_t$, $N_r$, $d(K_p,N_t,N_r)$ (since $\sum_{k\in\mathcal{K}_p}S_k\le d(K_p,N_t,N_r$), 
and the product $\rho_{k,k}\|\bH_{k,k}\|_F$. Knowledge about the number of antennas $N_t$ and $N_r$ is assumed known a priori, and the degrees-of-freedom depends
on the transmission strategies available~\cite{YetJafKay:Feasibility-Conditions-for-Interference:09,GhaMotKha:Interference-Alignment-for-the:09}, which are also known in advance.
The channel quality metric $\rho_{k,k}\|\bH_{k,k}\|_F^2$ can be estimated from the previous channel realization since large scale fading, 
including path loss and shadowing, is likely to be correlated across channel realizations. If $\rho_{k,k}\|\bH_{k,k}\|_F^2$ is not known exactly, we can substitute 
$\mathbb{E}\rho_{k,k}\|\bH_{k,k}\|_F^2$
in its place, given previous channel measurements.
At the beginning of the algorithm, however, $K_p$, $p\in\{1,\dots,P\}$ is undefined because the number of groups $P$ are unknown. One could perform the greedy algorithm for each
possible $P$ and choose the one with the highest sum rate, but this would increase the computational complexity of the algorithm by a factor of $K$. We can instead intelligently
choose $P$ based solely on a priori knowledge of $N_t$, $N_r$, $T$, $\mathcal{L}(K,N_t,N_r)$, and $d(K,N_t,N_r)$.
In particular, we define \emph{degrees of freedom with overhead} $\tilde{d}_K(k,N_t,N_r,T)$ as
\begin{equation}
\tilde{d}_K(k,N_t,N_r,T) = \frac{T/\lceil K/k\rceil -\mathcal{L}(k,N_t,N_r)}{T}d(k,N_t,N_r).
\end{equation}
We then choose 
\begin{equation}
K_O=\arg\max_{k\in\mathbb{N}_1}\tilde{d}_K(k,N_t,N_r,T)
\label{eq:k_o}
\end{equation}
and set $P=\left\lceil\frac{K}{K_O}\right\rceil$.
This choice of $P$ will be near a good overhead-capacity tradeoff since $K_O$ is the DOF-optimal number of users in an $N_t\times N_r$ 
interference channel with overhead $\mathcal{L}(k,N_t,N_r)$ and coherence time $T$.


Once $P$ is found, we can assign users to each group by their approximate rate $R_{k,p}$.
The algorithm is summarized in Table~\ref{tab:algo}.
\begin{table}
\centering
\begin{tabular}{|rl|}
\hline
1. & Find $K_O$ according to~(\ref{eq:k_o})\\
2. & $P=\lceil\frac{K}{K_O}\rceil$\\
3. & Set $\mathcal{K}_A=\{1,\dots,K\}$ and $\mathcal{K}_p=\emptyset$ for $p\in\{1,\dots,P\}$\\
4. & Find $\overline{R}_{k,p}$ for $k\in\mathcal{K}_A$ and $p\in\{1,\dots,P\}$\\
5. & Let $\{k',p'\}=\arg\max_{k,p} \overline{R}_{k,p}$\\
6. & Add $k'$ to the set $\mathcal{K}_{p'}$ and remove from $\mathcal{K}_A$\\
7. & If $\mathcal{K}_A\ne\emptyset$, return to 4; else done\\
\hline
\end{tabular}
\caption{Greedy algorithm based on IA rate and group size approximations.}
\label{tab:algo}
\end{table}
The algorithm in Table~\ref{tab:algo} requires $P(\sum_{i=0}^{K-1}K-i)$ searches, which grows approximately with $K^3$ assuming $P$ grows linearly with $K$ ($P$ will not grow
faster than linearly with $K$, so this is a worst-case analysis). Further, relative to the optimal search, this algorithm does not require computation of precoders (which
may be an iterative procedure for $K>3$), and does not require any of the channel coefficients to be trained and estimated.
Note that this algorithm is based on a model with linear precoding, which does not result
in a linear relationship between $K$ and $d(K,N_t,N_r)$~\cite{YetJafKay:Feasibility-Conditions-for-Interference:09,YetGouJaf:On-Feasibility-of-Interference:10}. This algorithm 
can work for non-linear precoding~\cite{GhaMotKha:Interference-Alignment-for-the:09}, which may
increase the degrees of freedom in a constant-coefficient interference channel, with an appropriate approximation of $\overline{R}_{k,p}$.
That problem is beyond the scope of this paper.

Finally, because this is a greedy algorithm, the addition of a user to the network is straightforward and efficient. One need only run the algorithm for the new user, with
a search complexity of $P$. After several users have joined the network, it will need to be restructured (likely with higher $P$), but for an incremental change, network
topology does not need to change. When a user leaves the network, the network can be maintained by re-allocating the user with the worst performance in the network. This keeps
the groups balanced without having to restructure at every change. Detailed exploration of this matter is left for other work~\cite{NosAndHea:User-admission-in-MIMO:11}.

\subsection{Sum Rate Fairness}\label{subsec:fairness}
The algorithm of Section~\ref{subsec:max_rate} allotted an equal amount of time in the frame for each group and maximized the sum rate under this constraint.
Maximizing the sum rate with unbalanced time allocation will lead to the group with highest sum rate transmitting for the entire frame. Unbalanced time allocation can be used, however,
to provide each group with the same sum rate. A disadvantage of such a design is that the group with the lowest sum rate is invariably using most of the frame. To mitigate such a 
scenario, we must carefully assign users to groups.

We first define the estimated sum throughput of group $p$ at any point in the algorithm to be 
$\overline{R}_p = \sum_{k\in\mathcal{K}_p}\overline{R}_{k,p}$.
We then define \emph{network disparity} for a particular allocation of users as
\begin{equation}
\rho(\{\overline{R}_p\})=\max_{\hat{p}\ne\hat{q}}\overline{R}_{\hat{p}} - \overline{R}_{\hat{q}}.
\end{equation}
We then modify Step~5 of the algorithm in Table~\ref{tab:algo} to be
\begin{equation}
\{k',p'\}=\arg\min_{k,p}\rho(\{\overline{R}_1,\dots,\overline{R}_p+\overline{R}_{k,p},\dots,\overline{R}_P\}).
\end{equation}
This modified algorithm will attempt to allocate sum rate equally among all groups. The rate, in general, will not be equal even after this algorithm modification, so group
transmission times must be allocated unequally. This allocation can be done based on the estimated sum rates $\{\overline{R}_p\}$ or the actual sum rates $\{R_p\}$ if
performed after all the training, estimation, and feedback for the frame has occurred. For simplicity we will use $\{R_p\}$. 
If group $p$ is allocated $\mu_pT$ symbols for transmission (including overhead), then the sum rate of the network becomes
\begin{equation}
R_{\rm sum} = \sum_{p=1}^P\frac{\mu_pT-\mathcal{L}(K_p,N_t,N_r)}{T}R_p
\end{equation}
We constrain $\sum_p\mu_p=1$ and $\mu_p\ge0,\forall p$. The sum rate of each group is an unknown $R^*$. We can enforce the equal-rate constraint with a set of equations:
\begin{equation}
\mu_pR_p - R^* = \alpha_pR_p,~p\in\{1,\dots,P-1\},\\
\end{equation}
which we can then form into a linear relation
\begin{equation}
\left(\begin{array}{cccccc} 
1      & 1      & 1      & \cdots & 1      &  0\\ 
R_1    & 0      & 0      & \cdots & 0      & -1\\
0      & R_2    & 0      & \cdots & 0      & -1\\
\vdots & \vdots & \ddots & \ddots & \vdots & \vdots
\end{array}\right)\left(\begin{array}{c} \mu_1\\ \mu_2\\ \vdots\\ R^* \end{array}\right) = \left(\begin{array}{c} 1\\ \alpha_1R_1\\ \vdots\\ \alpha_PR_P\end{array}\right),
\label{eq:time_solve}
\end{equation}
The time allocation vector $\mathbf{\mu} = \left(\mu_1, \mu_2, \dots, \mu_P, R^*\right)^T$ has a unique solution since the left matrix in~(\ref{eq:time_solve}) is square and non-singular.

\subsection{Geographic Grouping}\label{subsec:spatial}
Since the greedy Balanced Time Allocation algorithm proposed in Section~\ref{subsec:max_rate} estimates its rate based only on the SNR between user pairs~$\rho_{k,k}$, and neglects
inter-user SNRs $\rho_{k,\ell}, k\ne\ell$, it does not take advantage of possible natural groupings that may arise from geographical clusters. 
It has been shown that IA performs best, relative to other transmission techniques, when all receivers have strong links
to all transmitters. This is because IA is a degrees-of-freedom-optimal transmission strategy, and degrees-of-freedom are most important in the regime where
all receivers have strong links to all transmitters. Thus, a position- or signal strength-based algorithm could group geographically close users to maximize 
the benefit of IA.
Conversely, if non-IA transmissions are considered, a similar algorithm could group together users that are geographically separated, choosing to transmit
as if no interference existed. Since this regime is not ``high SNR'' in the interference channel sense (some links may have strong power, but not all), interference
alignment is not the desired transmission strategy, and instead interference can be ignored.
This section analyzes the latter case, which, as we will show, is algorithmically equivalent to the first case. 

We study the problem of geographic grouping under time-orthogonal transmissions, still considering the overhead model of previous sections.
It is assumed that the central controller executing the partitioning algorithm has position information for each transmitter and receiver in the network, although this could be
replaced with a channel quality indicator for the channels between all receivers and transmitters in the network. The position of receiver $k$ is $\mathbf{\delta}_k$, while the
position of transmitter $\ell$ is $\mathbf{\pi}_\ell$, so that the distance between transmitter $\ell$ and receiver $k$ is $\|\mathbf{\delta}_k-\mathbf{\pi}_\ell\|$. We then define
\begin{equation}
\Delta_{k,p}=\min_{\ell\in\mathcal{K}_p}\|\mathbf{\delta}_k-\mathbf{\pi}_\ell\|.
\label{eq:spatial_min}
\end{equation}
If no user is allocated to the $p$th group, then we define $\Delta_0>0$ to be a small default distance. 
Then $\Delta_{k,p}=\Delta_0$.
We can then modify Step~5 of the algorithm in Table~\ref{tab:algo} to be
\begin{equation}
\{k',p'\} = \arg\max_{k,p}\Delta_{k,p}.
\label{eq:spatial_max}
\end{equation}
To group the closest users and perform IA, we can simply switch the $\max$ and $\min$ 
in~(\ref{eq:spatial_min}) and~(\ref{eq:spatial_max}).

\section{Optimizing Training Overhead with Partitioning}
\label{subsec:imperfect_csi}
To analyze the relationship between channel partitioning and overhead, we consider the physical layer overhead of training for
channel estimation. While different interference alignment techniques have varying requirements for transmit CSI (and thus feedback overhead),
they all require receive CSI for interference nulling. The obtainment of receiver CSI is typically performed through transmission of a 
known training sequence orthogonal to the data. In this section we find the optimal training lengths for a given partition, and the effect
that partitioning has on training length.

In general, channel estimation is done in the presence of noise, which means imperfect CSI at both receiver and transmitter. In this case,~(\ref{eq:partition_rate}) is no longer
achievable. To approach this problem, we first assume perfect feedback of the imperfect channel estimates $\{\hat{\bH}_{k,\ell}\}$. Second, each receiver applies an 
interference-cancelling orthogonal filter $\bU_k$ to its received signal $\by_k$, such that $\bz_k = \bU_k^*\by_k$,
and $\bs_k$ is estimated using ML detection from $\bz_k$. Finally, we assume that the precoder design and receiver designs treat the channel state knowledge as perfect.

Let 
\begin{equation}
\hat{\bH}_{k,\ell} = \bH_{k,\ell} - \bE_{k,\ell}, \forall k,\ell.
\end{equation}
If the receivers use a minimum mean square error (MMSE) estimator, and the channel is being estimated in additive white
Gaussian noise, then $\hat{\bH}_{k,\ell}$ and $\bE_{k,\ell}$ are uncorrelated.
The precoder $\hat{\bF}_\ell$ is based solely on $\{\hat{\bH}_{k,\ell}\}$.
The received signal is then
\begin{equation}
\by_k = \sqrt{\rho_{k,k}}\bH_{k,k}\hat{\bF}_k\bs_k + \sum_{\ell\ne k}\sqrt{\rho_{k,\ell}}\bH_{k,\ell}\hat{\bF}_\ell\bs_\ell + \bv_k,
\end{equation}
and the filtered signal, after interference filtering, is
\begin{equation}
\bz_k = \sqrt{\rho_{k,k}}\hat{\bU}_k^*\hat{\bH}_{k,k}\hat{\bF}_k\bs_k + \sqrt{\rho_{k,k}}\hat{\bU}_k^*\bE_{k,k}\hat{\bF}_k\bs_k + \hat{\bU}_k^*\sum_{\ell\ne k}\sqrt{\rho_{k,\ell}}\bE_{k,\ell}\hat{\bF}_\ell\bs_\ell + \hat{\bU}_k^*\bv_k,
\label{eq:filtered_sig}
\end{equation}
since $\hat{\bU}_k^*\sum_{\ell\ne k}\hat{\bH}_{k,\ell}\hat{\bF}_\ell={\bf 0}$ through interference alignment. 
If the error matrix $\bE_{k,\ell}$ is drawn from a circularly symmetric complex Gaussian distribution, where each component is independent with variance $\sigma^2_E$
$\forall k,\ell$, previous work~\cite{TreGui:Cellular-interference-alignment:09} has found a lower bound on the sum rate using interference alignment with linear 
precoders when all the links have equal channel estimation error:
\begin{equation}
R_{EE} \ge \frac{T-\mathcal{L}(K,N_r,N_t)}{T}\sum_{k=1}^K\log\left|\frac{1}{\sigma^2_ES_k\gamma_{k,k}+1}\left(\bI\left(1+\sigma^2_E\gamma_{k,k} + 
            \sum_{\ell\ne k}\sigma_E^2\gamma_{k,\ell}\right) + \rho_{k,k}\tilde{\bH}_{k,k}\tilde{\bH}_{k,k}^*\right)\right|.
\end{equation}
Because of the homogenous assumption of channel estimation error, this formulation is useful when each receiver is roughly equidistant to each transmitter.
In general, such an assumption may not be valid. Further, by characterizing the error variance in terms of the training length, it is possible to design the length of our training
sequences to further trade overhead and rate.

Expanding the analysis to include unequal error variances as a function of training length, we extend a previous model for point-to-point 
communications~\cite{HasHoc:How-much-training-is-needed:03} to MIMO interference channels.
The residual interference term in~(\ref{eq:filtered_sig}) is possibly non-Gaussian and dependent on the data we wish to 
decode~\cite{HasHoc:How-much-training-is-needed:03}. We therefore find a lower-bound on the capacity of this
system by examining the worst-case additive noise that is uncorrelated with the data. This noise model is tractable and has the same energy as the residual interference term. 
The analysis from~\cite{HasHoc:How-much-training-is-needed:03} is directly applicable because we are making the same assumptions as Theorem~1 in that paper.
Thus, the worst-case uncorrelated additive noise is spatially white, zero mean,
circularly symmetric, and Gaussian with covariance matrix $\sigma^2_{\bn_k}\bI$. Refer to the Appendix of~\cite{HasHoc:How-much-training-is-needed:03} for proof. 
Define $\tilde{\bE}_{k,\ell}=\hat{\bU}_k^*\bE_{k,\ell}\hat{\bF}_\ell$. Assuming $\mathbb{E}\bs_\ell\bs_\ell=\bI,\forall\ell$, and 
$\mathbb{E}\bs_\ell\bs_k={\bf 0},\forall\ell\ne k$, then
\begin{eqnarray}
\sigma^2_{\bn_k} & = & \frac{1}{S_k}\mathrm{tr}\mathbb{E}\bn_k\bn_k^*\nonumber\\
{} & = & \frac{1}{S_k}\mathbb{E}\mathrm{tr}\left(\sum_{\ell=1}^K\frac{\rho_{k,\ell}}{S_\ell}\tilde{\bE}_{k,\ell}\tilde{\bE}_{k,\ell}^*\right)+1\nonumber\\
{} & = & 1 + \sum_{\ell=1}^K\frac{\rho_{k,\ell}}{S_\ell}\mathbb{E}\mathrm{tr}\left(\tilde{\bE}_{k,\ell}\tilde{\bE}_{k,\ell}^*\right)\nonumber\\
{} & = & 1 + \sum_{\ell=1}^K\rho_{k,\ell}\sigma^2_{\tilde{\bE}_{k,\ell}}.
\end{eqnarray}
Define $\sigma^2_{\tilde{\bH}_{k,k}}=\mathbb{E}\mathrm{tr}\tilde{\bH}_{k,k}\tilde{\bH}_{k,k}^*$, and by the orthogonality principle, 
$\sigma^2_{\tilde{\bH}_{k,k}}=1-\sigma^2_{\tilde{\bE}_{k,\ell}}$.
Finally, we normalize each channel such that $\overline{\bH}_{k,k} = \tilde{\bH}_{k,k}/\sigma^2_{\hat{\bH}_{k,k}}$.
The sum capacity with overhead is bounded from below by
\begin{equation}
R_\tau \ge \mathbb{E}\frac{T-\tau-\hat{\mathcal{L}}(K,N_t,N_r)}{T}\sum_{k=1}^K\log\left|\bI + 
       \frac{\rho_{k,k}\sigma^2_{\tilde{\bH}_{k,k}}}{1+\sum_{\ell=1}^K\rho_{k,\ell}\sigma^2_{\tilde{\bE}_{k,\ell}}}\frac{\overline{\bH}_{k,k}\overline{\bH}_{k,k}^*}{S_k}\right|,
\label{eq:imperfect_sum_rate_1}
\end{equation}
where $\tau$ is the number of transmissions per frame used for training, and $\hat{\mathcal{L}}(K,N_t,N_r)=\mathcal{L}(K,N_t,N_r)-\tau$ is the number of transmissions per 
frame required for overhead other than training. 

Utilizing orthogonal training sequences from each transmit antenna, we find that
\begin{eqnarray}
\sigma^2_{\tilde{\bH}_{k,k}} & = & \frac{\rho_{k,k}\tau}{S_k + \rho_{k,k}\tau}\\
\sigma^2_{\tilde{\bE}_{k,\ell}} & = & \frac{S_k}{S_k + \rho_{k,k}\tau}.
\end{eqnarray}
The sum rate~(\ref{eq:imperfect_sum_rate_1}) can thus be rewritten as
\begin{eqnarray}
R_\tau & \ge & \mathbb{E}\frac{T-\tau-\hat{\mathcal{L}}(K,N_t,N_r)}{T}\sum_{k=1}^K\log\left|\bI +
       \rho_{{\rm eff},k}\frac{\overline{\bH}_{k,k}\overline{\bH}_{k,k}^*}{S_k}\right|\\
{} & = & \mathbb{E}\frac{T-\tau-\hat{\mathcal{L}}(K,N_t,N_r)}{T}\sum_{k=1}^K\log\left|\bI +
       \frac{\rho_{k,k}^2\tau}{S_k + \rho_{k,k}\tau + \sum_{\ell=1}^K\frac{S_k+\rho_{k,k}\tau}{S_\ell+\rho_{k,\ell}\tau}\rho_{k,\ell}S_\ell}
       \frac{\overline{\bH}_{k,k}\overline{\bH}_{k,k}^*}{S_k}\right|.
\label{eq:imperfect_sum_rate_2}
\end{eqnarray}
The training length $\tau$ can then be found by Monte Carlo methods to maximize the lower bound of~(\ref{eq:imperfect_sum_rate_2}). 

At high SNR, 
\begin{equation}
\rho_{{\rm eff},k}\approx\frac{\rho_{k,k}\tau}{\tau+\sum_{\ell=1}^KS_\ell},
\end{equation}
and the effect of the residual interferers is constant with respect to $\{\rho_{k,\ell}\}$, meaning that there is no reduction in the degrees of freedom region compared to the perfect
CSI case. If $\{\rho_{k,\ell}\}$ is fixed, however, and $K$ increases, then sum throughput is reduced. 
Thus, to maintain a sum rate increase with additional users, the signal power must increase
with the addition of each user. Increasing the training length can also improve $\rho_{{\rm eff},k}$,
but is detrimental to the pre-log overhead factor.

If the network is partitioned into $P$ groups, each utilizing IA, then the rate is bounded from below by
\begin{equation}
\hat{R}_\tau \ge \mathbb{E}\sum_{p=1}^P\left(\frac{T/P - \tau_p - \hat{\mathcal{L}}(K_p,N_t,N_r)}{T}\right)\sum_{k\in\mathcal{K}_p}\log\left|\bI + 
                 \rho_{{\rm eff},k,p}\frac{\overline{\bH}_{k,k}\overline{\bH}_{k,k}^*}{S_k}\right|,
\label{eq:partitioned_training}
\end{equation}
where
\begin{equation}
\rho_{{\rm eff},k,p} =  \frac{\rho_{k,k}^2\tau_p}{S_k + \rho_{k,k}\tau_p + \sum_{\ell\in\mathcal{K}_p}\frac{S_k+\rho_{k,k}\tau_p}{S_\ell+\rho_{k,\ell}\tau_p}\rho_{k,\ell}S_\ell},
\end{equation}
and $\tau_p$ is the number of training symbols used in group $p$.
In this case, partitioning has the added potential to benefit the rate by reducing the total estimation error by reducing the number of channels to estimate.
The rate bound of~(\ref{eq:partitioned_training}) allows an engineer to design the length of the training sequences as functions of the expected channel conditions. In summary, partitioning an interference channel can not only increase throughput by reducing overhead, but it can also
increase the reliability of channel estimations. Further, the amount of overhead, $\alpha$, can be optimized through training length 
minimization. 

\section{Simulations}\label{sec:sims}
This section presents numerical results demonstrating the effect of overhead on the interference channel and comparing the greedy partitioning method of Section~\ref{sec:greedy}
to previous approaches.
The simulations are done using iterative interference alignment with linear 
precoding~\cite{GomCadJaf:Approaching-the-Capacity-of-Wireless:08,PetHea:Interference-Alignment-Via-Alternating:09} 
with 100 iterations, although the analysis does not preclude utilization of other IA designs. 
As in~\cite{PetHea:Cooperative-Algorithms-for-MIMO:09}, five random initializations are used at each iteration, and the 
precoding design with best sum throughput among the different initializations is chosen as the design for that iteration. 
The degrees of freedom using this method has been conjectured to be 
$d(K,N_t,N_r)=(N_t+N_r)K/(K+1)$~\cite{YetJafKay:Feasibility-Conditions-for-Interference:09} and the number of streams are varied according to this relationship. Thus, if 
a group has $K_p$ users with $M$ antennas, each transmitter will send $d(K_p,M)/K_p$ streams. When $d(K_p,M)/K_p$ is not an integer, some transmitters (chosen randomly) will transmit 
$\lceil d(K_p,M)/K_p\rceil$ streams and the rest will send $\lfloor d(K_p,M)/K_p\rfloor$ streams such that the sum of streams in the network is $d(K_p,M)$.
Unless noted otherwise, channels are generated with independent and identically distributed (i.i.d.) zero-mean circularly symmetric complex Gaussian coefficients with 
unit variance and $\rho_{k,\ell}=20$~dB, all 
$k$ and $\ell$, ensuring that the network is 
fully connected as discussed in Section~\ref{sec:model}. At low SNR values, interference alignment has been shown to perform poorly~\cite{GomCadJaf:Approaching-the-Capacity-of-Wireless:08},
thus the moderately high SNR environment is assumed. Generating the channels i.i.d.~with a Gaussian distribution gives an idea of the best possible performance, since
correlation has been shown to reduce IA rates~\cite{ElAPetHea:The-feasibility-of-interference-alignment:10}.
Absolute values for coherence time and overhead are irrelevant, so the overhead percentage of the coherence time, 
$\alpha=\mathcal{L}(K,N_t,N_r)/T$, or the data percentage of the coherence time assuming $P=1$ group, and $\overline{\alpha}=(T-\mathcal{L}(K,N_t,N_r))/T$, are used. 
For TDMA, overhead is assumed to scale linearly with the number of users ($\mathcal{L}=K$), while for IA, it scales with the square 
of the number of users~\cite{ThuBol:Interference-alignment-with:09} ($\mathcal{L}=K^2$).

\begin{figure}
\centering
\includegraphics[width=5in]{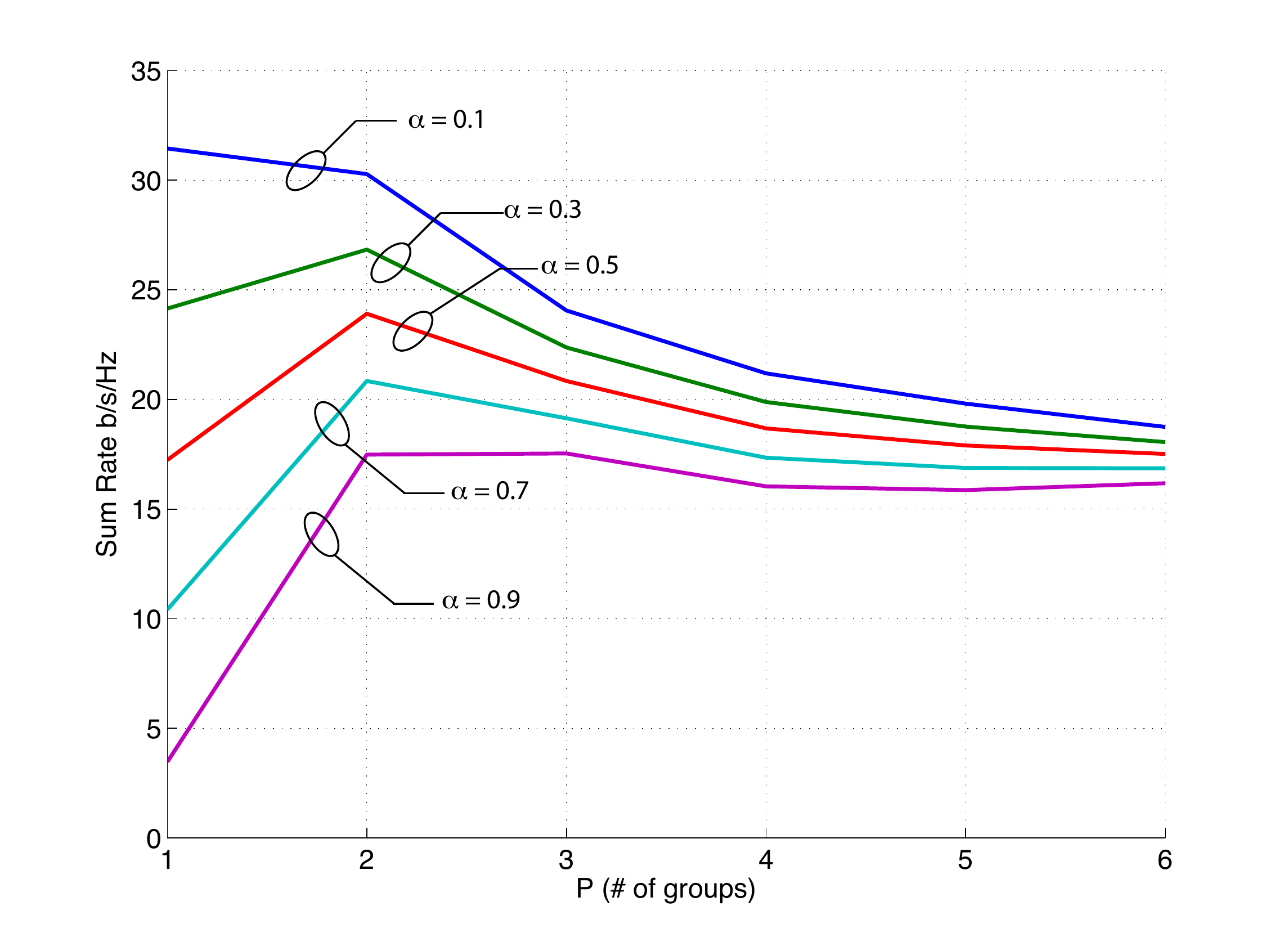}
\caption{Sum rate versus number of groups for the 6-user MIMO interference channel with $N_t=3$ and $N_r=4$. 
In this figure, $P=1$ groups corresponds to IA over the entire network while $P=6$ groups corresponds to TDMA.
With large $\alpha$, IA is not practical because overhead dominates the frame. As the coherence time increases ($\alpha$ decreases), however, $P=1$ (i.e., applying 
IA over the network) is the sum-rate-optimal partition.}
\label{fig:partition_scaling}
\end{figure}

Figure~\ref{fig:partition_scaling} demonstrates how the optimal number of groups in a partition of a 6-user network varies with the coherence time of the channel. In this figure,
$P=1$ means IA over the entire network, while $P=6$ means TDMA over the entire network. Thus, $P$ can be viewed as a complexity parameter that can vary transmission complexity
from IA to TDMA with every combination thereof, as depicted in Figure~\ref{fig:spectrum}. With low
coherence times (or high overhead percentage since overhead is constant for variable $T$), IA over the entire network results in overhead 
consuming the entire frame. TDMA gives a non-zero sum rate but is still not optimal. Partitioning the network into 4 groups, 2 of which have one user while the rest have two users,
results in the highest sum rate when overhead is considered. As the coherence time increases, however, IA gains start to outweigh the cost of overhead and thus a single-group partition,
equivalent to not partitioning the network, is the best choice in terms of sum rate.

\begin{figure}
\centering
\includegraphics[width=5in]{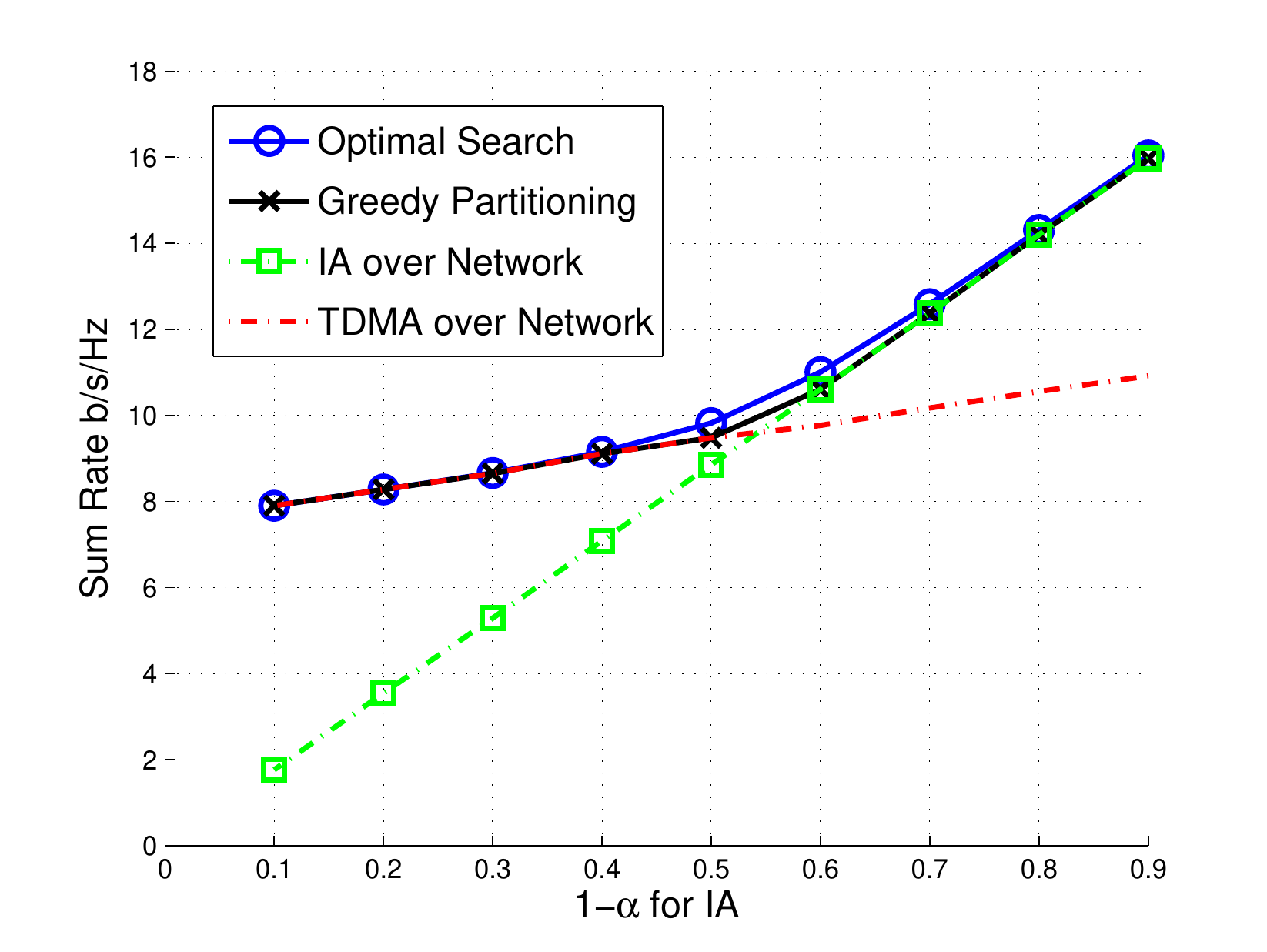}
\caption{Sum rate versus $\overline{\alpha}$ for exhaustive search, greedy partitioning, IA, and TDMA. For this simulation, the users are kept at $K=3$
and there are $N_t=N_r=2$ antennas at each node, one stream is sent by each transmitter in groups utilizing IA, and 2 streams are sent when a group consists of one node. 
The horizontal axis corresponds to the percentage of the coherence interval available for data transmission after overhead.
At low coherence times, the overhead required for IA dominates its performance and utilizing TDMA results in a better sum rate. As the coherence time increases,
IA gains begin to outweigh the overhead costs, and IA has a higher sum rate.} 
\label{fig:interval}
\end{figure}

\begin{figure}
\centering
\includegraphics[width=5in]{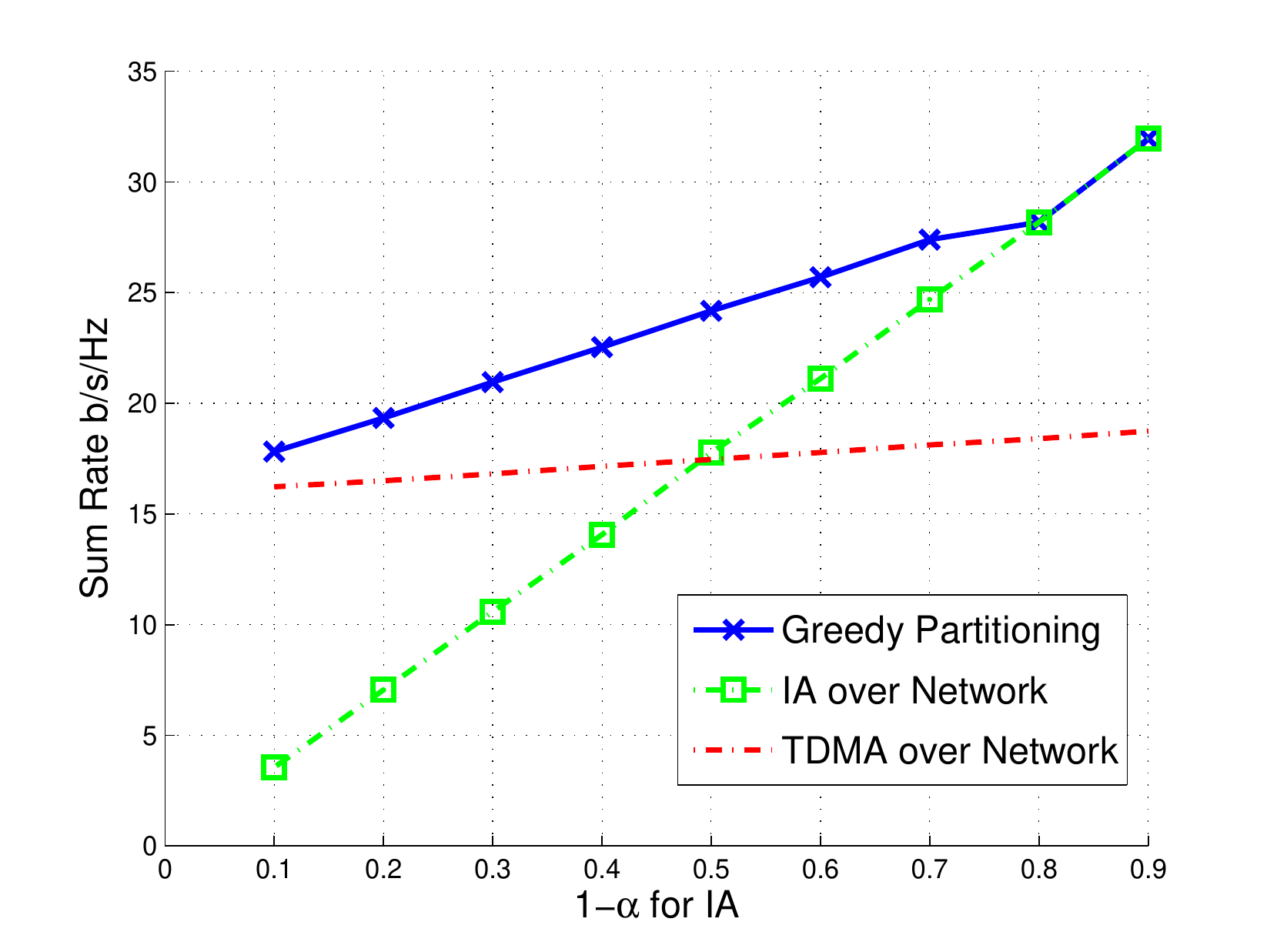}
\caption{Sum rate versus $\overline{\alpha}$ for greedy partitioning, IA, and TDMA. For this simulation, the users are kept at $K=6$
and there are $N_t=3$ antennas at each transmitter and $N_r=4$ antennas at each receiver. 
As in Figure~\ref{fig:interval}, the horizontal axis corresponds to the percentage of the coherence interval available for data transmission after overhead.
A larger gain is available when partitioning with more users relative to the 3 users of Figure~\ref{fig:interval}.}
\label{fig:interval_6users}
\end{figure}

\begin{figure}
\centering
\includegraphics[width=5in]{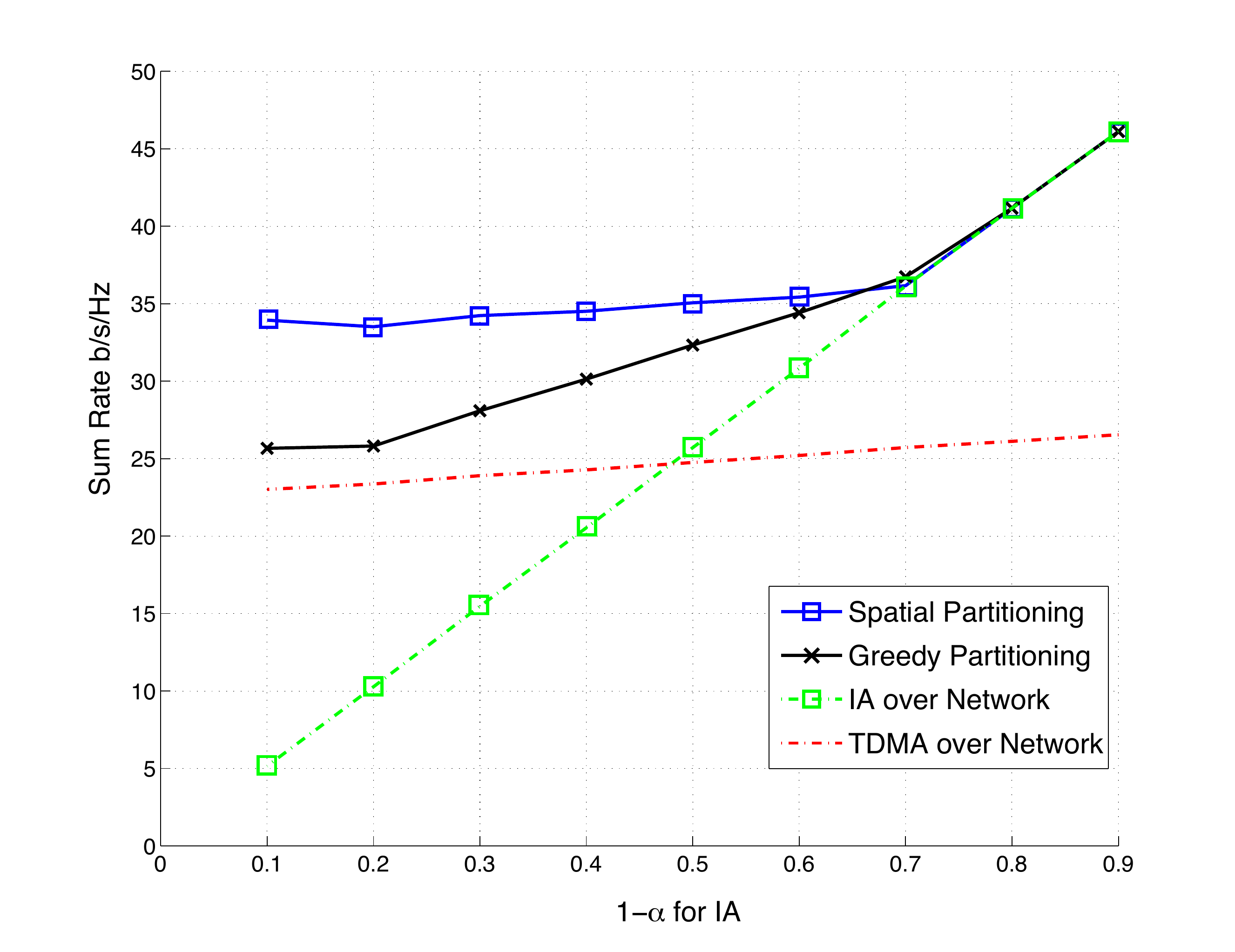}
\caption{Sum rate versus $\overline{\alpha}$ for greedy partitioning with geographic grouping, IA-sum
rate grouping, IA, and TDMA. For this simulation, a cellular channel model is used for a 6-cell
arrangement.}
\label{fig:spatial_cellular}
\end{figure}

Figure~\ref{fig:interval} shows the sum rate of the greedy partitioning method and the exhaustive partitioning method 
for $K=3$ users for various $\overline{\alpha}$, with $N_t=N_r=2$ antennas are at each node. For exhaustive partitioning, all possible values for $P$
are considered and the actual sum rate with global channel knowledge~(\ref{eq:partition_rate}) is used. With a small coherence time, TDMA outperforms IA,
whereas with a large coherence time, IA throughput gains outweigh the overhead cost of implementation, resulting in better sum rate than TDMA.
The partitioning algorithms are able to dynamically vary the network transmission strategy as the coherence time changes. Further, the greedy partitioning 
method, approximates the optimal partitioning without a brute force search, with its worst performance at moderately low $\overline{\alpha}$ due to the 
a priori choice of the number of groups based on degrees of freedom with overhead.
For $K=6$ users, partitioning leads to a larger sum rate increase at moderate SNR versus switching between IA and TDMA, as shown in Figure~\ref{fig:interval_6users}. 
This is due to the increased number of possible partitions. Greedy partitioning is again able to adapt between the possible partitions as overhead is varied. Note that,
although optimal search is not shown in this figure due to computational complexity, we know that since the greedy algorithm performs the partitioning based on large
scale statistics, its throughput curve as a function of $1-\alpha$ is a piecewise linear function. The different segments of this function are points where a particular 
partition size is judged to be favorable when averaged over small scale fading effects. This is visible in Figures~\ref{fig:interval},~\ref{fig:interval_6users}, 
and~\ref{fig:spatial_cellular}. Thus, the greedy algorithm will be furthest from optimal in the switching regions, such as around $1-\alpha\approx0.5$ in Figure~\ref{fig:interval}.
The gap between optimal and greedy will therefore grow with the number of possible partitions, and thus the number of users.

Figure~\ref{fig:spatial_cellular} demonstrates the gains of geographic grouping in a 6-cell network
with user locations drawn uniformly from a circle with radius 758~m around each base station,
which are placed 1.52~km apart. 
The channel
model is the Type~E model from IEEE~802.16j~\cite{Senoth:Multi-hop-Relay-System:07}, 
and the base stations transmit with $N_t=3$
transmit antennas and $40$~dBm transmit power. When the partitioning algorithm chooses $P>1$,
grouping the users based on geographic distance
outperforms the IA max-sum-rate algorithm because the IA gains are smaller in this operating
region and are offset by the relatively high overhead of IA versus ignoring the interference. 
That is, users can be grouped to operate in a high SIR region, where ignoring interference is preferable to aligning it. More spatial streams
can be exploited this way, utilizing less overhead because fewer channels must be estimated and fed back. 
At large coherence times IA is still the preferred strategy because the transmitters can utilize the entire frame after overhead for transmission. 

\begin{figure}
\centering
\includegraphics[width=5in]{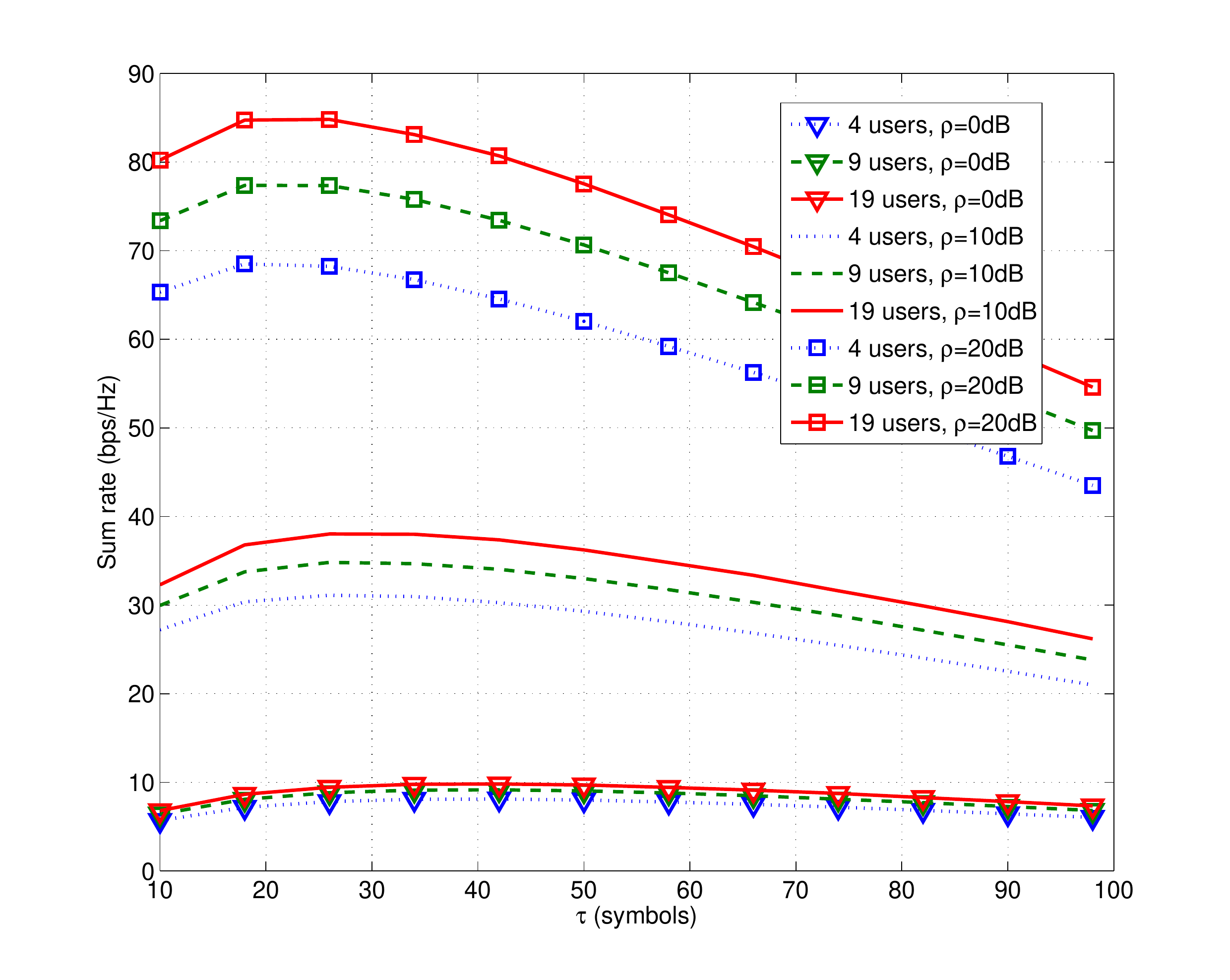}
\caption{Sum rate lower bound versus $\tau$ for $K=4,9,19$ users, $M=10$ antennas, $\rho=0,10,20$ dB,
and coherence time $T=200$ symbols. Optimal $\tau$ values for $\rho=0,10,20$ dB are 18, 26, and 42
symbols, respectively.} 
\label{fig:imperfect_csi}
\end{figure}

Finally, Figure~\ref{fig:imperfect_csi} demonstrates the lower bound on the sum capacity
from Section~\ref{subsec:imperfect_csi} as a function of the training length $\tau$ for
$M=10$ antennas, $K=4,9,19$ users, $\rho=0,10,20$ dB on all links, and coherence time $T=200$ symbols. 
In this case, the optimal
$\tau$ does not significantly vary for different $K$, but increases from $18$ to $42$ symbols
as $\rho$ decreases from $20$ dB to $0$ dB.

\section{Conclusions}\label{sec:conclusion}
This paper demonstrated the limitations of cooperative protocols for interference channels through overhead that scales faster than linearly
with the number of users in the network. In particular, as the network grows, the sum rate with overhead of interference alignment goes to zero.
By considering network overhead in the practical design for the interference channel, this paper has found analytical and algorithmic methods for 
trading off the overhead with the sum rate increase of cooperative transmission strategies by partitioning the network into orthogonally transmitting groups. 
A suite of transmission designs spanning the simplicity of TDMA to the performance of IA can be chosen using the simple algorithms derived in this paper. 
The proposed algorithms attempt to maximize the sum rate with overhead with fair time sharing of the channel, fair sum rate between groups, or geographic
grouping to exploit the reduced interference levels in unconnected channels. More work is required to characterize and reduce the overhead required for
such strategies, particularly for obtaining CSI at the transmitters.




\bibliographystyle{IEEEtran}
\bibliography{IEEEabrv,../../bibliographies/relay_coordination/relay_coordination,../huawei_consult/trunk/Multicell,../../bibliographies/mimo_relay/mimo_relay,../../bibliographies/mimo_basics/mimo_basics,../../bibliographies/interference_channel/interference_channel,../../bibliographies/information_theory/information_theory,../../bibliographies/ia_convergence/ia_convergence}

\end{document}